\documentstyle[prl,aps]{revtex}
\begin{document}
\draft
\title{Nonextensive statistics in stellar plasma 
and solar neutrinos}

\author{A. Lavagno and P. Quarati}
\address{Dipartimento di Fisica, Politecnico di Torino, 
         C.so Duca degli Abruzzi 24, I-10129 Torino, Italy\\
Istituto Nazionale di Fisica Nucleare, Sezioni di Cagliari e di Torino}

\maketitle

\begin{abstract}
Nonextensive and quantum uncertainty effects (related to 
the quasiparticles composing the stellar core) 
have strong influence on the nuclear rates and, of course, affect solar 
neutrino fluxes. Both effects do coexist
and are due to the frequent collisions among the ions.
The weakly nonextensive nature of the solar core is confirmed.  
The range of predictions for the neutrino fluxes is enlarged and 
the solar neutrino problem becomes less dramatic. 
\vspace{1pc}
\end{abstract}

The stellar (like the solar) core is a weakly-nonideal plasma where: 
1) mean Coulomb energy potential is not much smaller of the thermal 
kinetic energy; 2) Debye screening length $R_D\approx a$ (interparticle 
distance) and Debye-H\"uckel conditions are only approximately 
verified; 3) it is not possible to separate individual and 
collective degrees of freedom; 4) inverse solar plasma 
frequency: $t_{pl}=\omega^{-1}_{pl}=
\sqrt{m/4\pi n e^2}\approx 10^{-17}$ 
is of the same order of magnitude of the collision 
time $t_{coll}=\nu^{-1}=\langle n \sigma v\rangle$; 5) particles 
of plasma loose memory of the initial state only after 
many collisions, scattering process cannot be considered Markovian; 
6) time needed to build up again the screening, after hard collisions, 
is not at all negligible \cite{kania97,corra}.

The relation between energy and momentum implied by 
$\delta(\epsilon-p^2/2m)$ for free particles is no more valid. 
In dense media like stellar core plasma, for quasi-particles, we must 
use \cite{kada,gali}
\begin{equation}
\delta_g(\epsilon)=\frac{1}{\pi} 
\frac{g(\epsilon,p)}{[(\epsilon-\epsilon_p-
\Delta(\epsilon,\epsilon_p))^2+g(\epsilon,\epsilon_p)^2]}\; ,
\end{equation}
where $\epsilon_p=p^2/2m$, 
$\Delta(\epsilon,\epsilon_p)$ and $g(\epsilon,p)$ are the real 
and ima\-gi\-na\-ry parts of the one-particle retarded Green's 
function self-energy. 

For weakly non ideal plasmas one can show that approximately 
$\Delta\approx kT\, \Gamma/2$, where $\Gamma$ is the pla\-sma parameter 
($\Gamma=e^2/R_D kT$) and $g \propto \hbar\nu$. At non zero value of $g$, 
a nonexponential tail appears in the distribution function $f_Q(p)$.  

In fact, for large momenta, we have \cite{gali,staro1,staro2}
\begin{eqnarray}
f_Q(p)&=&\int_{-\infty}^\infty \!\!\! f(\epsilon,p)\; \delta_g(\epsilon)\;  
 d\epsilon
\nonumber \\
&=&f_M(p)+\frac{1}{\pi}\int_{-\infty}^{[{\rm min}(\mu,\beta^{-1}]} 
\!\!\!\!\!\!\!\!\!\!  d\epsilon \;  
\frac{g(\epsilon,p)}{(\epsilon-\epsilon_p)^2}
\nonumber\\
&=&f_M(p)+\frac{h\nu}{2\pi} \,\frac{kT}{\epsilon_p^2} \,e^{\mu/kT} \; ,
\end{eqnarray}
$\mu$ being the chemical potential and $f_M(p)$ the Maxwellian distribution.

The value of the collision frequency $\nu$ is responsible of two different
 effects producing at high momenta important (although small) deviations 
of the  Maxwellian distribution $f_M(p)$:
\begin{itemize}
\item[$Q$)]
Quantum uncertainty effect: because of the frequent collisions, $f_M(p)$ 
can acquire a non-Maxwellian tail;
\item[$q$)]
Weak nonextensivity effect described by Tsal\-lis statistics with 
entropic parameter $q$ \cite{tsa} due to long-range interactions 
and non-Markovian memory effects; when deviation is small ($q\approx 1$) 
the distribution can acquire an enhanced or a depleted tail, the 
correction of $f_M(p)$ being given by the factor 
$\exp\left [-\frac{1-q}{2}\left(\frac{\epsilon_p}{kT}\right)^2\right]$ 
\cite{kania97,corra}.
\end{itemize}

Nuclear rates can be evaluated averaging the quasi-classical cross 
section $\sigma(\epsilon)$ over the momentum distribution, rather 
than the energy distribution, once we have substituted $\epsilon$ 
with $\epsilon_p$ \cite{staro2}. A rigorous derivation of reaction rates 
within the $Q$ effect can be found in \cite{savche}.

Deviations from the Maxwellian tail due to $Q$ and $q$ effects may 
lead to a strong increase or decrease of the nuclear rates in the 
solar core (solar models and solar neutrino problem are described 
in Ref.s \cite{brum1,brum2,caste}). $Q$ correction depends on the 
di\-stri\-bu\-tion $f(\epsilon,p)$ (Maxwell, Fermi, Tsallis,$\cdots$), 
on the collision cross section $\sigma(\epsilon)$ and collision 
frequency $\nu$. Nuclear tunnelling rates are 
\begin{equation}
k_{ij}^{Q}= k_{ij}^M \, (1+r_{ij}) \; ,
\end{equation}
where $r_{ij}$ is the  
ratio of the average $Q$ power law part respect to 
the Maxwellian part. The results are \cite{staro2}
\begin{eqnarray}
&&(pp): \ \ r_{11} (0.0759 R_\odot)=3.5\,\cdot 10^{-3} \; ;
\nonumber\\
\nonumber\\
&&(^3He,^3He): \ \ r_{33}=4.5 \,\cdot 10^8 \; ; 
\nonumber \\ 
\nonumber\\
&&(^3He,^4He): \ \ r_{34}=3 \,\cdot 10^9 \; ;
\nonumber\\
\nonumber\\
&&\frac{n_3}{n_3^M}\approx 10^8; \ \ 
 \frac{n_{Be^7}}{n^M_{Be^7}}=\frac{\Phi(Be^7)}
 {\Phi^M(Be^7)}=\frac{1}{50}\; ; 
\nonumber \\
\nonumber\\
&& \frac{n_8}{n_8^M}\approx 10^4 \nonumber
\end{eqnarray}
$L_\odot$ is conserved; neutrino pp 
flux $\Phi(pp)$ is unchanged.

When long-range interactions are present and detailed correlations 
in space and time exist it is no longer true that the probability of 
a particle being in a state and the probability of a transition 
are statistically independent thus the two probabilities 
cannot be multiplied. 
The natural generalization of the Boltzmann-Gibbs statistics is 
the Tsallis nonextensive thermostatistics that we have already applied  
elsewhere \cite{kania97,corra}. 

Based on the generalized entropy form
\begin{equation}
S_q=k \frac{1-\sum_i p_i^q}{q-1} \; \ \ \ 
\left (\sum_i p_i=1 \; , \ \ q \; {\rm real}\right ) \; ,
\end{equation}
Tsallis statistics uses conditional probabilities that, 
when $q<1$ and $q>1$, will respectively privilegiate the rare and the frequent events.

Among many applications, 
let us mention that from COBE data we have derived the distribution of 
peculiar velocities of clusters of spiral galaxies obtaining a 
remarkable fit (the function used was the $q$-generalized Maxwellian 
distribution 
essentially corresponding to an ideal classical gas) \cite{lava}.

In the solar core the random electric total microfield can be 
decomposed in three main components: 1) slow varying, due to 
collective pla\-sma oscillations, the particles see it as an 
almost constant external mean field over several collisions; 
2) fast random, described by elastic diffusive cross 
section ($\sigma\approx1/v$), the distribution re\-mains Maxwellian; 
3) short range two-body strong Coulomb effective interaction 
described by the ion sphere model with strict en\-for\-ce\-ment. 
The last is the component of our interest whose energy density 
can be expressed as $\langle {\cal{E}}^2\rangle=(F e/a^2)^2$. 
We have found that the following two relations hold:  
$F\simeq \alpha_1^{-2}$ and $F^2\simeq 3/\Gamma=40$.

The strong Coulomb cross section is $\sigma_0=
2\pi\alpha_1^2 \,a^2$ \cite{ichi}. The quantity $\alpha_1$ 
depends on the ion-ion correlation function. We have found the 
analytical relation
\begin{eqnarray}
|1-q|=\frac{2}{3} \frac{\sigma_0^2}{\sigma_1^2}=
12 \alpha_1^4\Gamma^2\ll 1\; . 
\end{eqnarray}
The quantity $\alpha_1$ may be defined as
\begin{eqnarray}
&&\alpha_1 a=\int_0^\infty \!\!\!P_{NN} (R)\, R \, dR=
\nonumber\\
&&\int_0^\infty \!\!\! dR\int_0^\tau \!\!\! dt \; 
R^3 4 \pi n_i g(R,t)  \times 
\nonumber \\
&&\exp\left ( -4\pi n_i \int_0^R \!\!\! dR^{'} 
\int_0^\tau \!\!\!dt \; R^{'2} g(R^{'},t) \right ) ,
\end{eqnarray}
where $P_{NN}(R)$ is the probability that nearest neighbor of ion $i$ is at a 
distance $R$ and 
$g(R,t)$ is the correlation function (eventually time dependent). 
The value of $\alpha_1$ is within the range 
$0.4<\alpha_1<0.89$.

The nuclear rates can be written 
\begin{equation}
k_{ij}^q=k_{ij}^M \; \exp(-(1-q)/2\, \gamma) \; ,
\end{equation}
with $\gamma=(E_{\rm Gamow}/4kT)^{2/3}$. \\
 
Using $q=0.99$ (for all particles) we obtain: \\

\noindent
$\Phi(pp)=62.2\cdot 10^9$ cm$^{-2}$ $s^{-1}$,  \\
\\
$\Phi(^7Be)=2.87\cdot 10^9$ cm$^{-2}$ $s^{-1}$, \\
\\
$\Phi(N,O)=0.21\cdot 10^9$ cm$^{-2}$ $s^{-1}$, \\
\\
$\Phi(B)=1.65\cdot 10^6$ cm$^{-2}$ $s^{-1}$ \\
(SSM: $5.15 \cdot 10^6$, exper.: $2.45\pm 0.08\cdot 10^6$), \\
\\
Gallium$=100$ SNU \\
(SSM: $129\pm 7$, exper. (Gallex): $77.5\pm 7.7$), \\
\\
Chlorine$=2.84$ SNU \\
(SSM: $7.7\pm 1.1$, exper.: $2.65\pm 0.23$). 

The quantum uncertainty $Q$ and nonextensive $q$ effects produce 
corrections to the standard nuclear rates that can be expressed as 
\begin{eqnarray}
k_{ij}=k_{ij}^M \Bigg [ \exp\left ( -\frac{1-q}{2}\,\gamma_{ij}\right) 
\nonumber \\
+r_{ij} \, \exp\left ( -\frac{1-q}{2}\,\gamma_{ij}^{*}\right)
\Bigg ]\; ,
\end{eqnarray}
both effects cannot be neglected due to the value of the collision frequency 
$\nu$.

The $Q$ effect is effective at higher momenta than $q$-effect 
$\gamma^{*}\simeq 3\gamma$.
Defining
\begin{eqnarray}
&&A=\frac{\Phi(Be^7)/\Phi^M(Be^7)}{\Phi(B)/\Phi^M(B)}\; , \ \ 
B=\frac{\Phi(B)}{\Phi^M(B)}\; , 
\nonumber \\
&&C=\frac{\Phi(Be^7)}{\Phi^M(Be^7)} \; ; \ \ \delta_{ij}=\frac{1-q_{ij}}{2}
\nonumber
\end{eqnarray}
(the index $M$ means that the flux is calcu\-la\-ted 
using the Maxwellian distribution) 
and $k_{e7}=x\, k_{17}^M \; , k_{e7}=y\, k_{17}$, we have derived the 
following constraints for the solar neutrino fluxes
\begin{eqnarray}
&&1)\; A=\frac{C}{B}=\frac{e^{\delta_{17}\gamma_{17}^{*}}}{r_{17}} \ \ 
e^{-\delta_{17}\gamma_{17}}\ll r_{17} e^{-\delta_{17}\gamma_{17}^{*}}\; , 
\nonumber \\
&&2)\; \frac{y}{x}\,\frac{1}{A}=1 \; , 
\nonumber \\
&&3)\; \frac{n_{Be^7}}{ n_{Be^7}^M}=\frac{n_3}{n_3^M}
 \left(e^{-\delta_{34}\,\gamma_{34}}
+r_{34} e^{-\delta_{34}\,\gamma_{34}^{*}}\right) 
\nonumber \\
&& \ \ \ \ \left [
1+\frac{1}{2x} \left (e^{-\delta_{17}\,\gamma_{17}}+r_{17}
 e^{-\delta_{17}\,\gamma_{17}^{*}} \right)
\right ]^{-1} \; . \nonumber
\end{eqnarray}

A reasonable evaluation of $\alpha_1$ gives $\alpha_1=0.55$; 
with $\Gamma=0.072$ (solar core) we have $q=0.989$ ($\delta_{ij}=0.005$) 
for all components. 

Assuming $n_3/n_3^M\simeq 3 \cdot 10^{-3}$, we obtain 
\begin{eqnarray}
&&\frac{\Phi(Be^7)}{\Phi^M(Be^7)}=\frac{1}{50}\; , \ \ \ 
\frac{\Phi(B)}{\Phi^M(B)}\approx 1/2 \; ,
\nonumber\\
&&{\rm Gallium}=81~{\rm SNU}\; , 
\nonumber \\
&&{\rm Chlorine}=2.8~{\rm SNU} \; ;
\nonumber
\end{eqnarray}
$\Phi(pp)$ and luminosity are 
practically unchanged respect to SSM value. The CNO reactions are 
strongly enhanced by the $Q$ effect but the $e^{-\delta\gamma^{*}}$ 
factor strongly reduces it. 

The assumption concerning the value of $n_3$ is within the constraints 
actually imposed by helioseismology because in the region $r/R_\odot<0.2$ 
the value of 
$n_3$ can be assumed within a large range of variability \cite{brum1,brum2}.\\

We thank S. Turck-Chi\`eze and V. Savchenko for comments and advices.

\end{document}